\documentclass[aps,prl,twocolumn,groupedaddress,showpacs,floatfix]{revtex4}
\usepackage{graphicx}
\usepackage{float}

\begin{document}

\title{Quantum phase transition in a single-molecule quantum dot}

\author{Nicolas Roch$^1$, Serge Florens$^1$, Vincent Bouchiat$^1$, 
Wolfgang Wernsdorfer$^1$ {\&} Franck Balestro$^1$}

\affiliation{$^1$Institut N\'eel, Nanosciences Department, associ\'e \`a l'UJF, CNRS, 
BP 166, 38042 Grenoble Cedex 9, France}

\begin{abstract}
Quantum criticality is the intriguing possibility offered by the laws of
quantum mechanics when the wave function of a many-particle physical system is
forced to evolve continuously between two distinct, competing ground states.
This phenomenon, often related to a zero-temperature magnetic phase
transition, can be observed in several strongly correlated materials such as
heavy fermion compounds or possibly high-temperature superconductors, and is
believed to govern many of their fascinating, yet still unexplained properties. 
In contrast to these bulk materials with very complex electronic structure,
artificial nanoscale devices could offer a new and simpler vista to the
comprehension of quantum phase transitions.
This long-sought possibility is demonstrated by our work in a fullerene molecular
junction, where gate voltage induces a crossing of singlet and triplet spin states
at zero magnetic field.
Electronic tunneling from metallic contacts into the $\rm{C_{60}}$ quantum 
dot provides here the necessary many-body correlations to observe a true 
quantum critical behavior.
\end{abstract}

\maketitle

\section{Introduction}

\begin{figure*}
 \includegraphics[width=17cm]{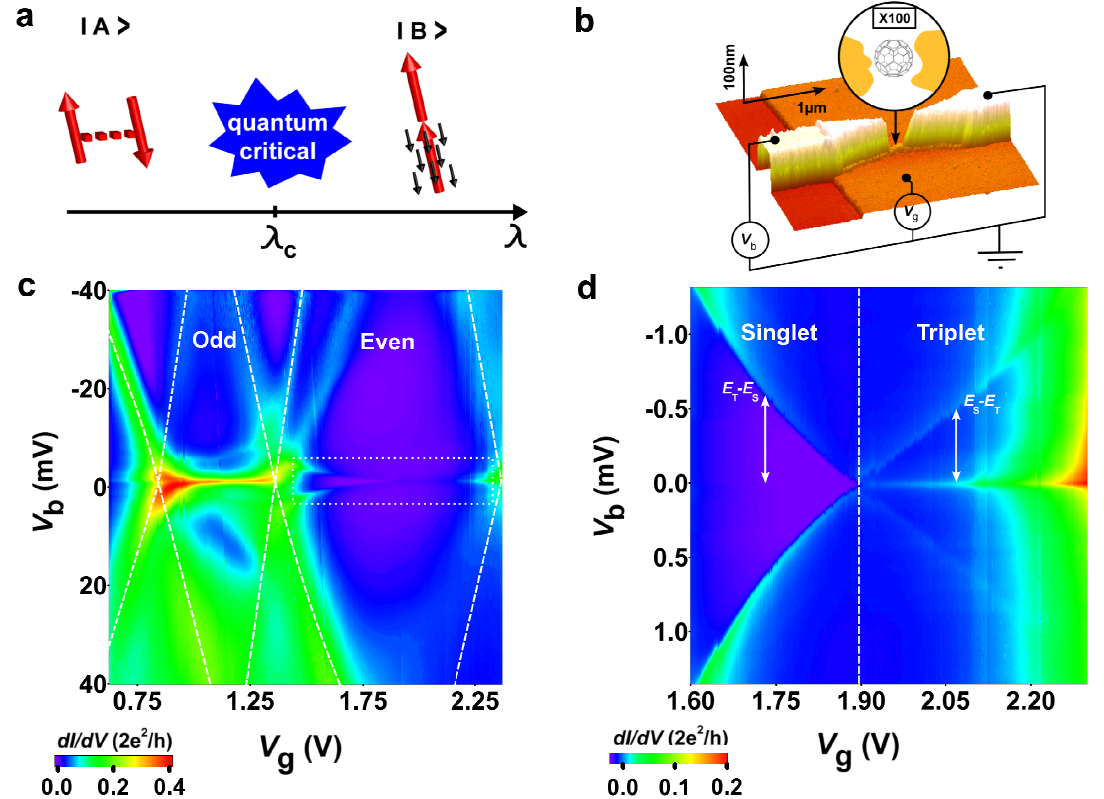}
 \caption{\label{fig1} {\bf Quantum phase transition, device and 
 device characteristics.}
 {\bf a,} Quantum phase transition: a quantum state $\left|A\right>$ can be driven by a non-thermal 
 external parameter $\lambda$ to another quantum state 
 $\left|B\right>$ with a different
 symmetry through a critical point at $\lambda=\lambda_\mathrm{c}$.
 In our $\rm{C_{60}}$ QD device, $\left|A\right>$ is a singlet, and $\left|B\right>$ is a 
 triplet, partially screened by one conduction electron channel represented
 by black arrows.
 {\bf b,} AFM photography of the device showing the gold nano-wire 
 over an $\rm{Al/Al_{2}0_{3}}$ gate. 
 A $\rm{C_{60}}$ molecule is trapped in the nanogap formed during the electromigration
 process.
 {\bf c,} Colour-scale map over two Coulomb diamonds of the differential conductance 
 $\partial I/\partial V$ as a function of bias voltage $V_{\rm{b}}$ and gate voltage 
 $V_{\rm{g}}$ at $35$~mK and zero magnetic field. 
 {\bf d,} Detailed characteristics of the differential conductance in the region ``Even'', corresponding to
 a low bias measurement inside the dotted white rectangle in panel 
 {\bf c}. This
 region marks the crossing of singlet and triplet spin states of the QD.}
 \end{figure*}

In order to grasp the fundamental difference between classical and quantum phase
transitions, one can start with the basic principle of thermodynamics that 
systems in thermal equilibrium lie in their minimal free energy state. 
As a consequence, a change in temperature can drive a classical phase 
transition by a competition between low energy configurations and the 
entropy associated to thermal fluctuations.
However, strictly at zero temperature, most physical systems have quenched their entropy 
and some kind of order may be expected. Even then, Heisenberg's uncertainty principle of 
quantum mechanics implies that zero-point motion is always present. If the strength of 
these quantum fluctuations can be increased by changing some non-thermal
parameter $\lambda$ (in analogy to an increase of temperature), the ground state of the system 
may evolve between two wavefunctions with different symmetries, $\left|A\right>$ and 
$\left|B\right>$, by a quantum phase transition at a critical value 
$\lambda\approx \lambda_\mathrm{c}$ (Fig.~\ref{fig1}a).
Interestingly, even if absolute zero temperature is experimentally unaccessible, quantum 
phase transitions leave peculiar fingerprints that are visible at non zero
temperature~\cite{Sachdev}.

Quantum Dots (QDs) seem then to be ideal devices to observe quantum phase transitions.
First, such gate-tunable artificial atoms offer a high degree of control by
simple gate electrostatics. Second, due to the nanometric confinement of the electrons, 
they display relatively high energy scales that allow the 
obsvervation of interesting quantum 
phenomena at low temperature. Finally, the coupling between the QD
and the electronic
reservoirs (transport probes) provides tunneling events 
 that can fundamentally 
alter the dot states into complicated many-body wavefunctions.
One well studied situation (although not classified as a quantum transition)
where these three points fortuitously play together occurs when a single unpaired spin $S=1/2$ 
characterizes the ground state of the QD. 
When conducting electrons move to-and-from the device while reversing the tiny 
magnetic moment of the dot, a progressive screening of the atomic spin occurs, in complete 
analogy to the well-known Kondo effect in solids containing magnetic 
impurities~\cite{Hewson,Glazman1988}. The Kondo effect in QDs is then observed as zero-bias 
conductance resonance~\cite{Goldhaber1998,Cronenwett1998}, associated to the entangled state of electrons in the
leads and in the dot, and which displays a high degree of
universality.

QDs with even occupancy introduce the possibility of having a singlet or a triplet spin state, 
if the coupling to the electrodes may be neglected. 
When the Kondo effect sets in to meet the singlet-triplet splitting, a subtle competition 
occurs for the fate of the dot magnetic state, which was predicted to turn the singlet-triplet 
crossing into a true zero-temperature quantum phase transition, although different theoretical
scenarios have been put forward, depending whether one~\cite{Vojta2002,Hofstetter2002} or 
two~\cite{Georges1999,Jones1988} screening channels are involved.
While quantum critical phenomena related to Kondo screening have been attributed to the 
astonishing properties of many strongly correlated materials~\cite{Loehneysen2007}, a 
clear-cut observation in nanostructures of a screening-unscreening transition 
is still lacking. 
This is despite the intensive studies of singlet-triplet Kondo effects measured 
through vertical quantum dots~\cite{Sasaki2000}, GaAs lateral quantum dots under an 
applied magnetic field~\cite{Schmid2000,Wiel2002}, or at zero magnetic 
field~\cite{Kogan2003,Vidan2006}, carbon nanotubes~\cite{Nygard2000,Jarillo2005,Paaske2006,Quay2007}, 
and double dot structures~\cite{Craig2004}. 
Indeed, the ability to observe a sharp quantum transition is limited either by the existence 
of two screening electronic channels (linked to conserved orbital quantum numbers), 
which generically give an avoided transition~\cite{Affleck1995,Zarand2006}, or by 
relatively low Kondo temperatures, leading to the broad features observed in those experiments. 

$\rm{C_{60}}$ QDs inserted in a nanoscale constriction present
several key ingredients for observing a quantum phase transition: i) due to their
tunneling geometry, a predominant single screening channel should be expected 
(similarly to lateral QDs); ii) previous investigations have 
demonstrated large Kondo
temperatures $T_{\rm{K}}$~\cite{Park2002,Ralph2002,Natelson2004} due to the small nanometer 
size of the QDs molecule; iii) as demonstrated below, gate-tuning of the 
singlet-triplet gap at zero magnetic field can be obtained and allows 
the precise tuning of the system across the singlet-triplet boundary.
This experiment constitutes an advance in the realization of many-body effects in
quantum dots. These results are thus relevant for understanding bulk correlated
materials and opens new possibilities for the precise control of spin states in
nanostructures containing few electrons.

\section{Experimental set up}

We briefly discuss the sample preparation and measurement system. A 
more detailed description details on the measurement is given in the
Supplementary Information. We emphasize
that the experiment was carried out in a dilution
refrigerator with a high degree of filtering, a technical development that
was crucial for unveiling the complex quantum phenomena that takes place at
low temperatures.
Preparation of the single $\rm{C_{60}}$ transistor was realized by blow drying 
a dilute toluene solution of the $\rm{C_{60}}$ molecule onto a gold nano-wire 
realized on an $\rm{Al/Al_{2}O_{3}}$ back gate, see Fig.~\ref{fig1}b for a
schematic view of the setup. Before blow 
drying the solution, the electrodes were cleaned with acetone, ethanol, 
isopropanol solution and oxygen plasma. The connected sample is 
inserted in a copper shielded box, 
enclosed in a high frequency low temperature filter, consisting of thermocoax 
micro-wave filter and $\Pi$ filters, anchored to the mixing chamber of 
the dilution fridge having a base temperature equals to $35$~mK.
The nano-wire coated with molecules is then broken by electromigration~\cite{McEuen1999}, 
via a voltage ramp at $4$~K. Our electromigration 
technique uses real time electronics
to monitor the coupling of the single molecule to the electrodes. 
Here we report a full experimental study of transport 
measurements on a $\rm{C_{60}}$ QD, as a function of bias voltage 
$V_{\rm{b}}$, 
gate voltage $V_{\rm{g}}$, temperature 
$T$~($35$~mK $<$ $T$ $<$ $20$~K), and magnetic field 
$B$ up to $8$~T.


\section{Overview of transport characteristics at zero magnetic field}


We start by discussing the general features of the $\rm{C}_{60}$ QD, on the basis of
a large scale two-dimensional map of the differential conductance 
$\partial I/\partial V$ as a function of both bias $V_{\rm{b}}$ and 
gate $V_{\rm{g}}$
voltages, obtained at $35$~mK and at zero magnetic field (Fig.~\ref{fig1}c). 
The distinct conducting and non-conducting regions are typical signatures of a single molecule transistor.
We present measurements
over two distinct Coulomb diamonds indicated by ``Odd'' and ``Even'' 
on Fig.~\ref{fig1}c.

In region ``Odd'', we measure a sharp high-conductance ridge in the zero-bias 
$\partial I/\partial V$, which is clearly associated to the usual $S=1/2$ Kondo 
effect~\cite{Goldhaber1998,Natelson2004}.
This conductance anomaly, connected to the complete quenching of the dot local moment by
the conduction electrons, was already investigated at length for other quantum dot
systems (see~\cite{Grobis2006} for a review), and we report for comparison in the 
Supplementary Information a detailled study of the conductance as a function of 
temperature and magnetic field.

Henceforth we focus on the ``Even'' Coulomb diamond where an increase 
in gate voltage results in 
an additional electron on the dot and thus an even number of total electron.
These two-electron states can be described by their total spin 
$S$ and spin projection $m$ and are noted $\left|S,m\right>$. The
ground state of the system can thus be either a spin singlet $\left|0,0\right>$
with energy $E_{\rm{S}}$, or a spin triplet with energy $E_{\rm{T}}$ described by the three
states $\{\left|1,1\right>,\left|1,0\right>,\left|1,-1\right>\}$, 
degenerated at
zero magnetic field, but split by the Zeeman effect, with an energy shift
$\Delta E_{\rm{T}} = m g\mu_{\rm{B}}B$ for each state $\left|1,m\right>$, where ${g\approx{2}}$ 
for a $\rm{C_{60}}$ molecule (See Supplementary Information).
Fig.~\ref{fig1}d presents a precise low bias $\partial I/\partial V$ measurement 
of the ``Even'' region inside the dotted rectangle of Fig.~\ref{fig1}c, which
clearly displays two distinct regions, which we associate by anticipation to
the singlet and triplet ground states respectively. The possibility of a
gate-tuning of the singlet-triplet splitting $E_{\rm{T}}-E_{\rm{S}}$ was demonstrated previously both
for lateral quantum dots~\cite{Kogan2003} and carbon nanotube junctions~\cite{Quay2007}.
The two levels cross at a critical gate voltage
$V_{\rm{g}}^\mathrm{c}\simeq1.9$~V.

In the ``Singlet'' region, a finite-bias conductance anomaly appears 
when $V_{\rm{b}}$
coincides with $E_{\rm{T}}-E_{\rm{S}}$, and is explained by a non-equilibrium Kondo effect involving 
excitations into the spin degenerate triplet. This effect was recently elucidated on a 
carbon nanotube quantum dot in the singlet state~\cite{Paaske2006}. 
As shown in the Supplementary Information, our results are consistent with this previous report. 

Increasing the gate voltage, and neglecting the electrode coupling 
which will be discussed in greater details below, the triplet becomes the ground state.
Two kinds of
resonances are then observed: i) a finite-bias $\partial I/\partial 
V$ anomaly, interpreted 
as a singlet-triplet non-equilibrium Kondo effect on the triplet side, that disperses
as $E_{\rm{S}}-E_{\rm{T}}$ in the ($V_{\rm{g}}$,$V_{\rm{b}})$ plane;
ii) a sharp zero-bias $\partial I/\partial V$ peak, related to the underscreened 
spin $S=1$ Kondo effect~\cite{Nozieres1980}, as indicated by the low value of the
conductance peak.

In order to precisely identify these spin states, and vindicate our
later scaling analysis near the singlet-triplet crossing point, we now present a
detailed magneto-transport investigation of the ``Even'' region.

\section{Identification of the magnetic states in the ``Even'' diamond}

\begin{figure*}
 \includegraphics[width=15cm]{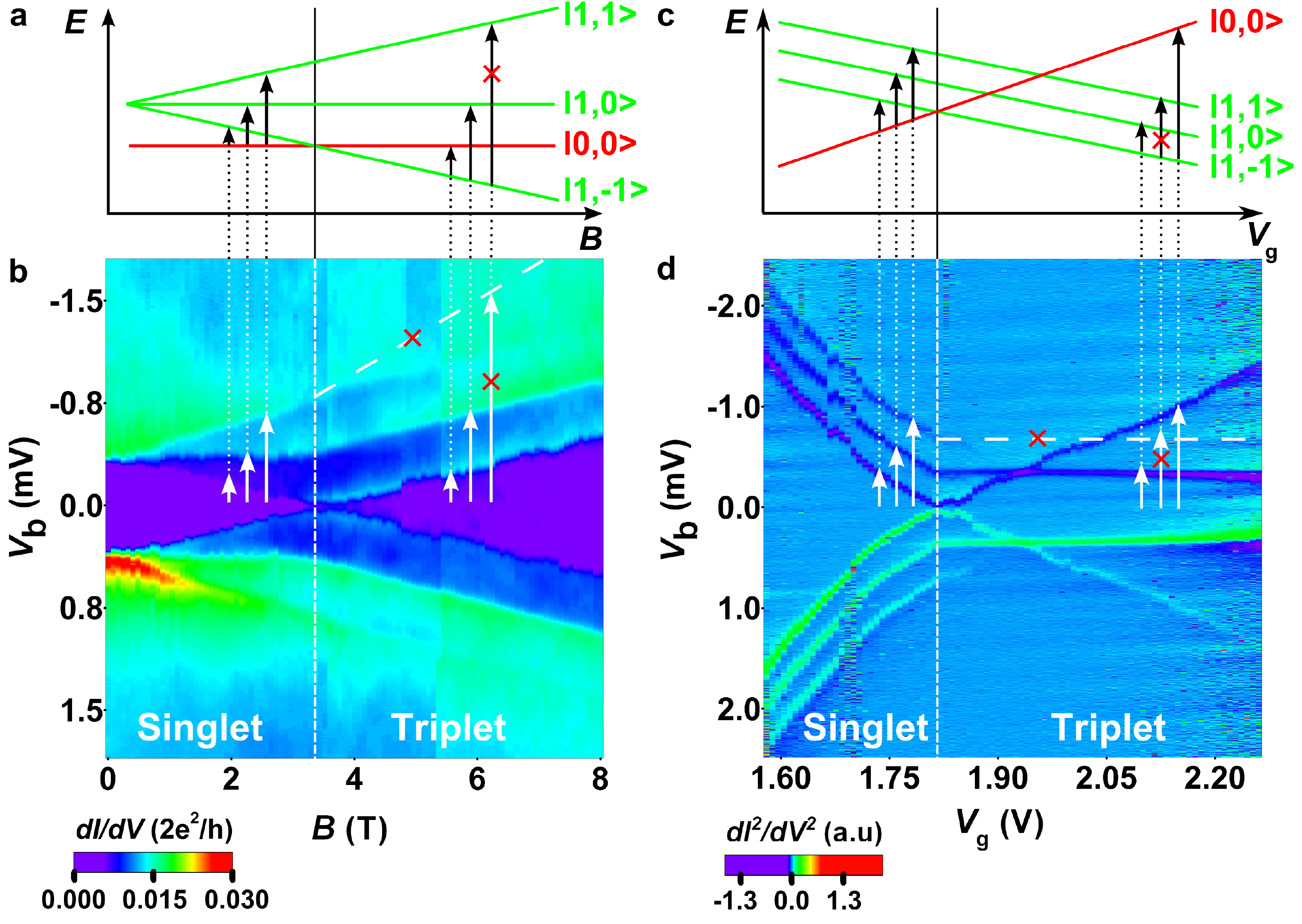}
 \caption{ {\bf Magnetic field and gate induced singlet-triplet 
 transition.} {\bf a,} Schematic of the singlet $\left|0,0\right>$ to lowest 
 triplet $\left|1,-1\right>$ transition induced by the Zeeman effect.
 {\bf b,} $\partial I/\partial V$ measurements as a function of $B$, at fixed gate
 voltage and temperature $T=\rm{35}$~mK. As symbolized by the
 crossed dotted line and arrow, second order spin flip processes with $\Delta{m}=2$ are not 
 observed, see also panel {\bf a}. 
 {\bf c,} Schematic of the singlet $\left|0,0\right>$ to lowest 
 triplet $\left|1,-1\right>$ transition induced by gate voltage at constant magnetic
 field.
 {\bf d,} $\partial^2 I/\partial V^2$ measurements as a function of the 
 gate voltage $V_{\rm{g}}$,
 at fixed magnetic field $B=3$~T and temperature $T=35$~mK.}
 \label{fig2}
\end{figure*}

Due to the high ${g\approx{2}}$ factor of a $\rm{C_{60}}$ molecule, 
as compared for instance to ${g\approx{0.44}}$ in GaAs-based devices, 
it is easier to lift 
the degeneracy of the triplet state via the Zeeman effect.
Figures \ref{fig2}b and \ref{fig2}d show the evolution of the different conductance
anomalies in the ``Even'' Coulomb diamond, as a funtion of magnetic 
field.

Fig.~\ref{fig2}b shows $\partial I/\partial V$ as a function of the 
magnetic field up to $8$~T, for a constant gate voltage $V_{\rm{g}}$ chosen in the
``Singlet'' region (albeit far from the transition point).
In this case, a Zeeman induced transition from the $\left|0,0\right>$ singlet to the 
lowest $\left|1,-1\right>$ triplet occurs by increasing the magnetic field, as
sketched in Fig.~\ref{fig2}a, and demonstrated by the clear level crossing in the
conductance map. The splitting of the threefold triplet is also blatant,
and the various spectroscopic lines are consistent with the spin selection rules
at both low and high field, where $\left|0,0\right>$ and $\left|1,-1\right>$ are
the respective ground states.

Fig.~\ref{fig2}d investigates the gate-induced singlet-triplet 
crossing at a constant magnetic field of 3~T. On the singlet side, the Zeeman splitted
triplet states are clearly seen as three parallel lines, while the transition
lines from the $\left|1,-1\right>$ ground state at higher voltage are in agreement with the
energy levels depicted in Fig.~\ref{fig2}c. Note that due to lower contrast in
the ``Triplet'' region, we have plotted here for better visibility the second order 
derivative of the current $\partial^{2} I/\partial^{2} V$.

This gate and magnetic field study gives thus proof of a singlet to triplet
transition inside the ``Even'' diamond. One further remarkable aspect of our
data is the absence of a large enhancement of the zero-bias conductance at the
singlet-triplet crossing, both in zero-field (Fig.~\ref{fig1}d) and in the
Zeeman effect (Fig.~\ref{fig2}b).
Such features were on the contrary observed in previous experiments. In vertical
semiconductor quantum dots~\cite{Sasaki2000}, a field-induced orbital effect can 
be used to bring the non-degenerate triplet in coincidence with the low-lying 
singlet state, leading to a spectacular Kondo enhancement of the conductance,
which was shown to be associated to two screening channels~\cite{Pustilnik2001}.
In carbon nanotube junctions~\cite{Nygard2000}, the Zeeman effect dominates
over the orbital effect, so that the transition involves the lowest triplet
state only, and Kondo signatures arise from a single channel, which must be however
well-balanced between the two orbital states of the quantum
dot~\cite{Pustilnik2000}.
The lack of both types of singlet-triplet Kondo effects points towards the
predominent coupling between a single screening channel and one of the
two spin states of the $\rm{C}_{60}$ QD, in which case a Kosterlitz-Thouless
quantum phase transition should infact be expected at the singlet-triplet 
crossing~\cite{Cornaglia,Hofstetter2002}.

\section{Investigation of the singlet-triplet quantum phase transition}
 
Having established the nature of the magnetic states far from the transition
point, we focus now on the conductance for gate voltages $V_{\rm{g}}$ close to the
critical value $V_{\rm{g}}^\mathrm{c}$ where singlet and triplet states are brought together at
zero magnetic field.

\begin{figure}
 \includegraphics[width=9.0cm]{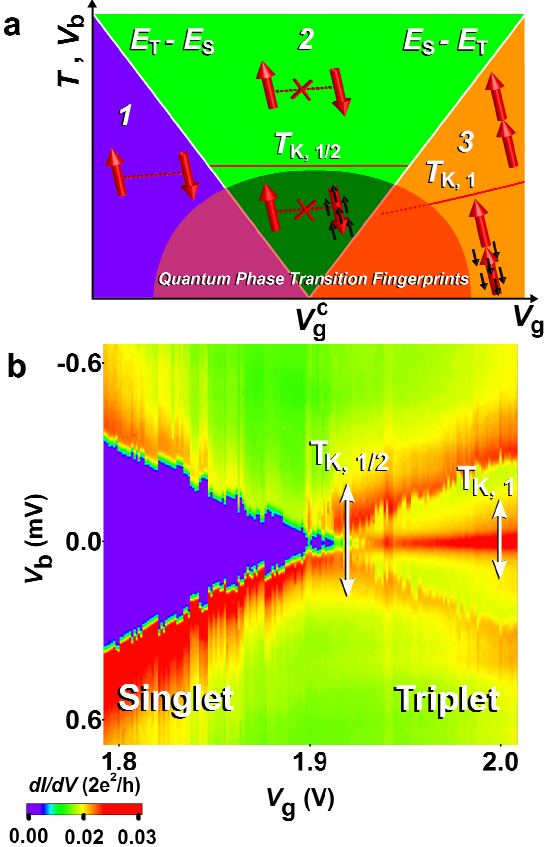}
 \caption{\label{fig3} {\bf Singlet-triplet quantum phase 
 transition.} {\bf a,} Phase diagram as a function of $V_{\rm{g}}$ : for simplicity 
 three differents regions can be identified depending whether 
 $V_{\rm{b}}$ (or
 $T$) lies above the singlet-triplet splitting $|E_{\rm{T}}-E_{\rm{S}}|$.
 The effective spin states of the QD are represented by large red arrows, and
 screening electrons by smaller grey arrows.
 The uncrossed dotted line between the spins signals a strongly bound singlet state 
 in region $\bf{1}$. In region $\bf{2}$ the two spin states decouple from each other,
 and the spin which is more strongly coupled to the leads is fully screened via
 a spin $1/2$ Kondo effect associated to the large Kondo temperature $T_{\rm{K},1/2}$.
 In region $\bf{3}$, the ground state of the QD is a $S=1$ triplet, and experiences
 an incomplete screening associated to the Kondo temperature $T_{\rm{K},1}$.
 {\bf b,} Colour-scale map of the differential conductance $\partial I/\partial V$ as a 
 function of bias $V_{\rm{b}}$ and gate $V_{\rm{g}}$ voltage at $35$~mK and zero magnetic 
 field, close to the singlet to triplet transition, where the scales
 $|E_{\rm{T}}-E_{\rm{S}}|$, $T_{\rm{K},1/2}$ and $T_{\rm{K},1}$ are clearly identified.}
\end{figure}

The transition region shown in the middle of Fig.~\ref{fig1}d is thus expanded
in the conductance map of Fig.~\ref{fig3}b taken at the base temperature $T=35$~mK. 
We first notice that a sharp conductance dip forms on the singlet side of the 
transition, in contrast to the shallow minima observed in previous experiments 
for two-level quantum dots in the singlet regime~\cite{Craig2004,Kogan2003,Quay2007}. 
Now focusing on the triplet side, we note the presence of two different energy
scales (that we will relate to two Kondo screening processes): 
i) the zero bias peak shows a small and narrowing width, which we connect to a
$S=1$ Kondo temperature $T_{\rm{K},1}$; ii) the satellites previously 
associated to a non-equilibrium singlet-triplet Kondo effect far from the transition 
point now smoothly merge together to form a large resonance related to another
larger Kondo temperature that we note $T_{\rm{K},1/2}$. Both Kondo scales, together with the 
singlet-triplet splitting, are reported on the schematic phase diagram of 
Fig.~\ref{fig3}a.
Several insights on the interpretation in terms of Kondo physics are given by the 
temperature dependence of the conductance for the different regions labelled.

\begin{figure*}
 \includegraphics[width=17cm]{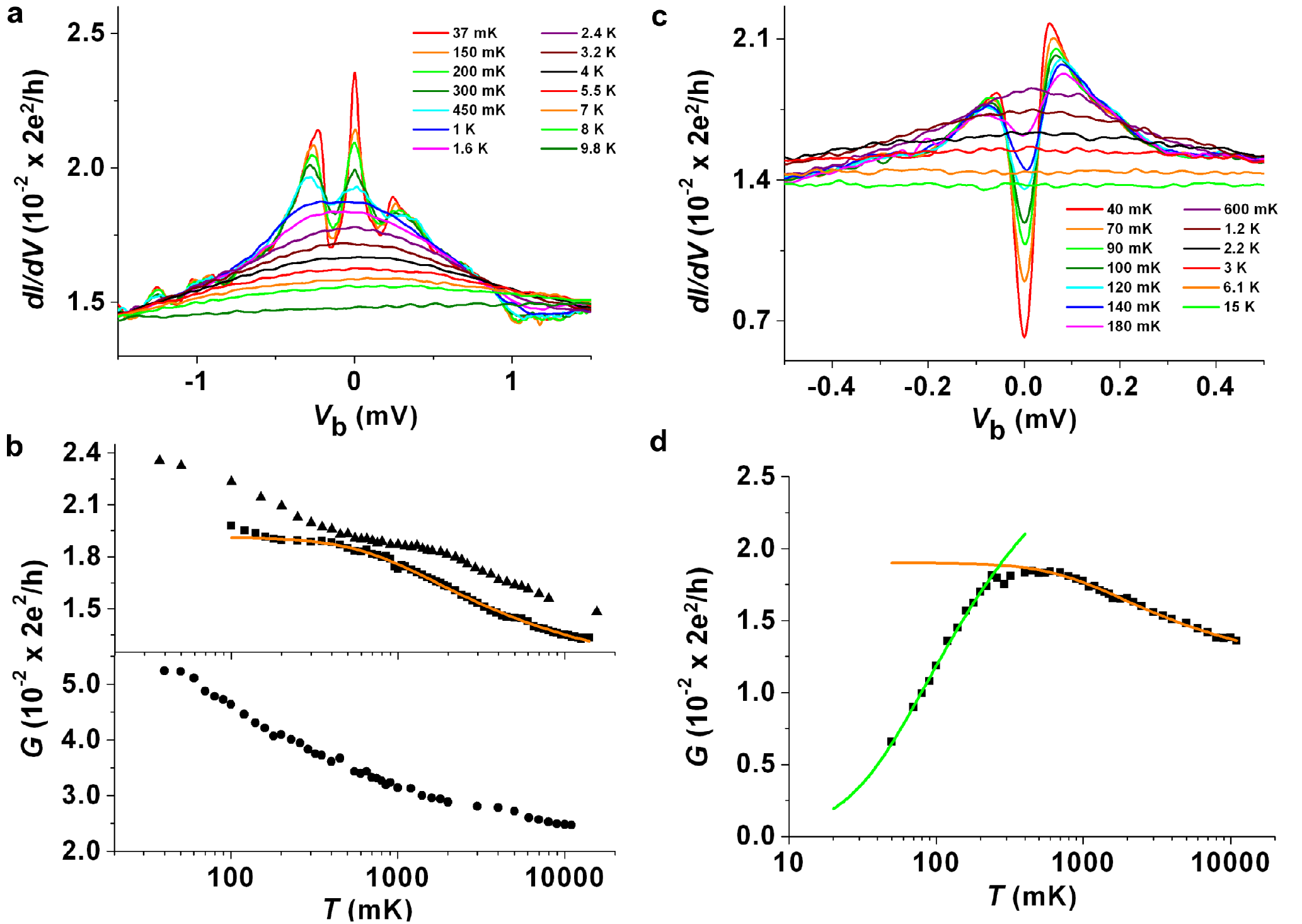}
 \caption{\label{fig4} {\bf Kondo effect in the singlet and triplet 
 state.}
{\bf a,} Differential conductance close to the transition point on the triplet side at 
different temperatures showing a broad resonance with Kondo temperature $T_{\rm{K},1/2}$ 
and Kondo satellites centered at $\pm|E_{\rm{S}}-E_{\rm{T}}|$.
{\bf b,} Temperature dependence of the zero bias conductance $G(T)$ for three gate
voltage values in the ``Triplet'' region, the topmost curve corresponding to the panel {\bf a}.
We clearly measure a gate 
dependent plateau, corresponding to the energy scale 
$\vert{E_{\rm{T}} - E_{\rm{S}}}\vert$. The solid line is a fit 
to Eq.~\ref{spin1/2} giving $T_{\rm{K,1/2}}=3.77$~K. The bottom 
panel does not show a clear plateau, and correspond to the 
temperature evolution of $G(T)$ in the underscreen spin $S=1$ regime.
{\bf c,} Differential conductance close to the transition point on the singlet side at 
different temperatures showing a broad resonance with Kondo temperature $T_{\rm{K},1/2}$ 
and a narrow dip~(inverse Kondo effect) associated with a temperature scale $T^\star$.
{\bf d,} Temperature dependence of the zero bias conductance $G(T)$ corresponding to
the panel {\bf c}. The orange line is a fit 
to Eq.~\ref{spin1/2} giving $T_{\rm{K,1/2}}=4.13$~K, and the 
green line is a fit to Eq.~\ref{invKondo} giving $T^{\star}=287$~mK.} 
\end{figure*}

\begin{figure}
 \includegraphics[width=9cm]{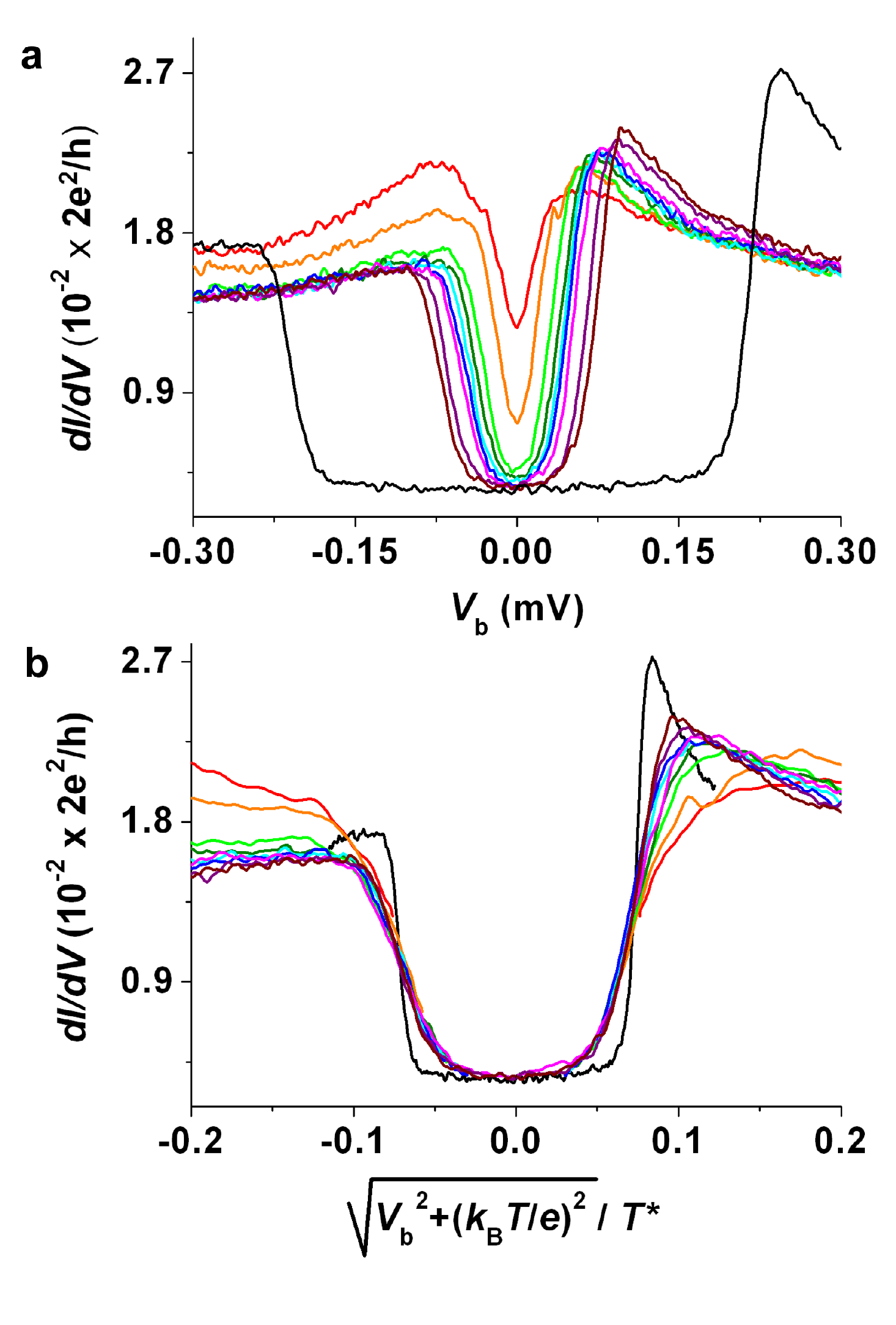}
 \caption{\label{fig5} {\bf Universal scaling.} {\bf a,} Differential conductance for different values of 
 gate voltage $V_{\rm{g}}<V_{\rm{g}}^\mathrm{c}$, close~(inverse 
 Kondo effect exhibiting a resonant dip) and far~(U-shaped curve) from the transition point.
 {\bf b,} Scaling analysis of the data of panel {\bf a}, with respect to the singlet
 binding energy $T^\star$.}
\end{figure}

In region $\bf{1}$ of Fig.~\ref{fig3}a we identify two different 
regimes. Far from the transition point, i.e. when the splitting 
$E_{\rm{T}}-E_{\rm{S}}$
exceeds the Kondo temperature $T_{\rm{K},1/2}$, the two spins are 
strongly coupled in 
a singlet state, and a non-equilibrium Kondo effect involving the degenerate 
excited triplet state is observed, as previously discussed. The differential 
conductance exhibits a characteristic U-shape associated with the singlet-triplet 
gap, as shown by the widest curve in Fig.~\ref{fig5}a. 
Close to the transition point, $E_{\rm{T}}-E_{\rm{S}}$ can now become 
smaller than $T_{\rm{K},1/2}$,
so that the formation of the singlet state involves the hybridization of the 
conduction electrons. The temperature
dependence of the differential conductance in Fig.~\ref{fig4}c shows the
formation of a narrow dip inside a broader resonance of width 
$k_{\rm{B}} T_{\rm{K},1/2}/e$.
This dip has been predicted theoretically~\cite{Hofstetter2002} and 
here is shown to behave as an inverted Kondo resonance. Specifically, 
Fig.~\ref{fig4}c shows that is has a Lorentzian line shape in bias voltage and
Fig.~\ref{fig4}d shows its logarithmic temperature dependence at zero bias. More quantative 
agreement with the theory can be made by fitting the temperature dependence of the 
conductance dip with the formula
%
%
\begin{equation}
G(T)=G_{0}\left[1-\left({\frac{T^{2}}{T^{\star 2}}\left(2^{1/s}-1\right)+1}\right)^{-s}\right]+G_{\rm{c}}
\label{invKondo}
\end{equation}
where the crossover scale $T^\star$ is related to the singlet binding
energy, $G_{0}$ is a typical conductance~(both are taken as free 
parameters), $G_{\rm{c}}$ is the background conductance and $s=0.22$.
Taking the experimental value of $G_{\rm{c}}$ far from the transition 
point, good agreement is found with this formula as shown
in Fig.~\ref{fig4}d. 
Since the formation of the singlet state is associated with an inverse Kondo effect 
with a characteristic temperature $T^\star$ that depends on the ratio 
$(E_{\rm{T}}-E_{\rm{S}})/T_{\rm{K},1/2}$, an universal behavior of the
conductance dip is expected near the transition point. 
Fig.~\ref{fig5}a shows that the differential conductance, at the base
temperature $T=35$~mK and for $V_{\rm{g}}<V_{\rm{g}}^\mathrm{c}$, 
evolves from a Lorentzian to a U-shape, and Fig.~\ref{fig5}b shows 
that this data can be rescaled~\cite{Goldhaber2006}  as a function of $\sqrt{V_{\rm{b}}^2+({k_{\rm{B}}} 
T/e)^2}/T^\star$. This latter plot shows that the conductance curves collapse on top of each
other when taken close to the transition point but that scaling deteriorates
as the singlet-triplet gap $E_{\rm{T}}-E_{\rm{S}}$ becomes greater 
than $T_{\rm{K},1/2}$.

The temperature dependence of $G(T)$ and the $V_{\rm{g}}$-scaling of 
$\partial I/\partial V$ provide strong evidence that the formation of the 
spin singlet state near the transition point involves a Kondo process at 
the low-temperature scale $T^\star$, which can be seen as a second stage of
screening. The binding of the singlet and its associated conductance dip 
(inverse Kondo peak) appear inside a much broader resonance of width 
$T_{\rm{K},1/2}$. We argue that this resonance is associated 
with the spin $S=1/2$ Kondo effect, the hallmark of the singlet-triplet quantum critical
point~\cite{Hofstetter2002}. To demonstrate this for $T>T^\star$ we 
fit in Fig.~\ref{fig4}d $G(T)$ using the 
formula~\cite{GoldhaberGordon1998}
%
%
\begin{equation}
   G(T)=G_{0}\left({\frac{T^{2}}{T_{\rm{K,1/2}}^{2}}\left(2^{1/s}-1\right)+1}\right)^{-s}+G_{\rm{c}}
\label{spin1/2}
\end{equation}
where $G_{0}$ is the conductance at $T=0$, $G_{\rm{c}}$ is the 
background conductance, and $s=0.22$.
The good agreement confirms our interpretation of the
critical domain (region $\bf{2}$ of Fig.~\ref{fig3}a) as a regular 
spin $S=1/2$
Kondo effect experienced by one of the two spins. The second spin is
disconnected from the leads in this energy range.

We finally turn to the triplet region $\bf{3}$ in Fig.~\ref{fig3}a. 
Far from the transition point, at large gate voltage $V_{\rm{g}}$ values, 
the spins are tightly bound into a triplet and a spin $S=1$ Kondo effect 
is expected.
Estimates from both the width of the zero-bias peak and its magnetic field
splitting (not shown) converge to a Kondo scale $T_{\rm{K},1}$ of the
order of $100$~mK. This value is unfortunately too low to allow quantitative
comparison with theoretical predictions of the underscreened Kondo effect. The
conductance $G(T)$ is shown on the lower panel in Fig.~\ref{fig4}b,
and does not show any sign of saturation down to our effective 
electronic temperature $T_{\rm eff}=50$~mK.
When the gate voltage is decreased, a complex regime, where the singlet-triplet
splitting $E_{\rm{S}}-E_{\rm{T}}$ is comparable to the high energy 
Kondo scale $T_{\rm{K},1/2}$, occurs.
This is shown by the differential conductance, at fixed $V_{\rm{g}}$ with lowering
temperature, in Fig.~\ref{fig4}a. 
While a broad peak is again observed at high temperatures, a three-peak structure 
emerges upon cooling. We interpret the latter by a non-equilibrium Kondo effect
that mixes singlet and triplet states via the voltage bias window.
We associate the former with a spin $S=1/2$ Kondo effect, similarly to what 
occurs on the singlet side. This idea is consistent with the corresponding zero-bias 
conductance $G(T)$ for temperatures above the singlet-triplet 
splitting, as shown by the upper panel in Fig.~\ref{fig4}b. 
The spin $S=1/2$ Kondo behavior of $G(T)$ extend down to the lowest 
temperatures by approaching the critical point, as $E_{\rm{S}}-E_{\rm{T}}$ becomes smaller 
than $T_{\rm{K},1/2}$. This is most clearly displayed by the lower curve of the same
plot, to which a fit with equation~(\ref{spin1/2}) can be successfully performed.
The further increase of $G(T)$ below 200mK is at present not fully understood, and 
may be related to the opening of a second screening channel, which would possibly
spoil the quantum critical point at zero temperature~\cite{HofstetterZarand}. 
However, this extra feature seems related to a very small energy scale close to 
the crossing point, so that the interpretation of the data in the accessible 
temperature range is consistent with the quantum critical point scenario.
We finally note that our data can be analyzed in a complementary way (given in 
the Supplementary Information) by plotting the zero-bias conductance as a fonction 
of gate voltage, for different temperatures. By cooling, this shows the clear 
sharpening of a conductance step when the system crosses from singlet to triplet, 
in agreement with the existence of a quantum critical point~\cite{Hofstetter2002}, 
and in contrast to the maximum predicted for an avoided 
transition~\cite{HofstetterZarand,Pustilnik2001}.

We end by noting that the singlet-triplet transition in a $\rm{C_{60}}$ molecular 
junction differs fundamentally from the observations in semiconductor~\cite{Sasaki2000}
and nanotube~\cite{Nygard2000} dots, thus showing the potentiality of combining 
well-controlled electromigration techniques with molecules of complex chemistry.

\begin{acknowledgments}
{\bf Acknowledgments} We gratefully acknowledge E. Eyraud, D. Lepoittevin for their useful 
electronic and dilution technical contributions and motivating 
discussions. We thank E. Bonet, T. Crozes and T. Fournier for lithography 
development, C. Winkelmann, T. Costi and L. Calvet for 
invaluable discussions. The sample of 
the investigations was fabricated in the NANOFAB facility of the 
N\'eel Institut.
This work is partially financed by ANR-PNANO Contract MolSpintronics No. 
ANR-06-NANO-27. 
\end{acknowledgments}

\end{document}